\documentclass[conference]{IEEEtran}
\usepackage{cite}
\usepackage{amsmath,amssymb,amsfonts}
\usepackage{algorithmic}
\usepackage{graphicx}
\usepackage{textcomp}
\usepackage{xcolor}
\usepackage{booktabs}
\usepackage{longtable}
\usepackage{pdflscape}
\usepackage{tabularx}
\usepackage{tikz}
\usetikzlibrary{shapes}
\usepackage{adjustbox}
\usepackage{array}
\usepackage{tikz}
\usetikzlibrary{arrows.meta,positioning,calc}
\usepackage{cuted}
\usetikzlibrary{positioning,shapes.geometric,}
\usetikzlibrary{shapes.geometric}
\usepackage{dblfloatfix}
\usepackage{caption}
\usepackage{url}
\usepackage[numbers,sort&compress]{natbib}
\usepackage[hidelinks]{hyperref}
\usepackage{ragged2e}
\newcolumntype{P}[1]{>{\RaggedRight\arraybackslash}p{#1}}

\def\BibTeX{{\rm B\kern-.05em{\sc i\kern-.025em b}\kern-.08em
    T\kern-.1667em\lower.7ex\hbox{E}\kern-.125emX}}
\begin{document}

\title{A Taxonomy of Real-World Asset Tokenization for Blockchain-Based Financial Infrastructure
\thanks{Omitted for a double blind review.}
}

\author{\IEEEauthorblockN{1\textsuperscript{st} Giorgio Vella}
\IEEEauthorblockA{\textit{School of Banking, Finance and Insurance} \\
\textit{Università Cattolica 
del Sacro Cuore
}\\
Milano, Italy \\
giorgio.vella01@icatt.it}

\and

\IEEEauthorblockN{2\textsuperscript{nd} Luca Pennella}
\IEEEauthorblockA{\textit{Interdisciplinary Centre for} \\
\textit{Security, Reliability and Trust} \\
\textit{University of Luxembourg}\\
Luxembourg, Luxembourg \\
luca.pennella@uni.lu}

\and

\IEEEauthorblockN{3\textsuperscript{rd} Mark C. Ballandies}
\IEEEauthorblockA{\textit{Blockchain and Distributed Ledgers} \\
\textit{University of Zurich}\\
Zurich, Switzerland \\
markchristopher.ballandies@uzh.ch }

}

\maketitle

\begin{abstract}
Real-world asset (RWA) tokenization has emerged as a prominent application of blockchain technology, enabling off-chain financial and non-financial assets to be represented through blockchain-based instruments. However, deployed RWA systems remain difficult to compare because legal claims, custody arrangements, token mechanics, verification processes, and on-chain integrations are often described separately. This paper develops a systems-level taxonomy of RWA tokenization to classify how off-chain assets are legally, economically, and technically represented on-chain. Following an iterative taxonomy-development method, we organize twenty-three dimensions into five components: governance, asset structure, token properties, distributed ledger technology, and economy. We apply the taxonomy to twenty major RWA systems selected by market capitalization and compare their design choices across asset classes and implementation models. The classification shows that current RWA tokenization is predominantly implemented through hybrid architectures: blockchain tokens support representation, transfer control, redemption workflows, pricing, and composability, while core legal guarantees remain anchored in off-chain legal wrappers, custodial arrangements, compliance processes, and verification mechanisms. The analysis also reveals recurring documentation gaps concerning voting rights, dispute forums, burn mechanics, supply constraints, and reserve verification. Overall, the taxonomy provides a structured basis for comparing RWA systems, identifying design patterns and limitations, and supporting future research on blockchain-based financial infrastructure.
\end{abstract}

\begin{IEEEkeywords}

Real-world assets, asset tokenization, RWA, decentralized finance, tokenized securities, smart contracts, taxonomy
\end{IEEEkeywords}

\section{Introduction}

Real-world asset (RWA) tokenization has emerged as an interdisciplinary research field at the intersection of distributed ledger technologies, smart contract engineering, financial economics, and legal and regulatory scholarship~\cite{luo2026sok}. By representing claims on assets through blockchain-based instruments, RWA systems introduce new forms of financial infrastructure in which asset representation, custody, settlement, compliance, and redemption mechanisms are partly encoded in technical architectures and partly governed by off-chain legal and institutional arrangements~\cite{lavayssiere2025legalstructures,bis2024tokenisationcpmi}.

Existing work, however, often examines these dimensions separately. Technical studies focus on token standards, settlement mechanisms, accounting models, oracle dependencies, and composability, whereas economic and policy-oriented analyses emphasize market frictions, institutional adoption, legal enforceability, and systemic risk~\cite{imf2025inefficiencies,fsb2024stabilitytokenisation}. This fragmentation limits comparability across RWA protocols, asset classes, and implementation models, and obscures how technical design choices interact with legal claims, governance structures, and market-level outcomes. Since RWA systems are interdependent complex systems~\cite{voshmgir2019foundations} spanning technical, legal, and socioeconomic dimensions, the field requires a shared analytical vocabulary to evaluate design choices and accumulate architectural knowledge~\cite{ballandies2022decrypting}.

As of May 2026, DeFiLlama tracks approximately $\$26\,\text{billion}$ in RWA market capitalization and more than 100 issuers across different asset classes, implementation models, functional scopes, and application domains \cite{defillama_rwa}. Yet, a systematic taxonomy jointly capturing the technical, legal, and market-design dimensions of RWA tokenization remains underdeveloped. Such a taxonomy can clarify protocol differences, reveal unexplored architectural configurations, and guide the evaluation and design of RWA implementations~\cite{ballandies2022decrypting}.

This paper provides an initial step toward a comprehensive classification of RWA tokenization systems for blockchain-based financial infrastructure. Building on existing taxonomies of blockchain systems~\cite{ballandies2022decrypting,ballandies2025general,ballandies2023taxonomy}, we adapt and extend this stream of literature to the RWA domain to bridge the lack of common terminology for comparing deployed systems. The main contributions are as follows:
\begin{itemize}
\item We develop an empirically applicable taxonomy of RWA tokenization systems composed of 23 attributes, organized across five components: governance, asset structure, token properties, distributed ledger technology (DLT), and economy.
\item We apply the taxonomy to twenty leading RWA protocols by market capitalization, showing how deployed systems combine off-chain legal structures, custodial arrangements, verification mechanisms, token mechanics, and on-chain integrations.
\item We identify recurring RWA design patterns and limitations, highlighting the prevalence of hybrid architectures in which blockchain-based instruments remain dependent on off-chain legal claims, compliance processes, custody arrangements, and reserve verification.
\end{itemize}

\section{Related Work}

From an economic and market-design perspective, tokenization is commonly described as a programmable redesign of the asset lifecycle, including issuance, trading, servicing, and redemption. Potential benefits include faster settlement, reduced coordination costs, and lower reliance on legacy intermediaries. However, macro-financial analyses emphasize that these benefits are conditional rather than automatic. Tokenized markets may introduce new sources of fragility, including smart-contract vulnerabilities, operational dependencies, platform concentration, oracle risk, and stronger interconnections across financial infrastructures~\cite{imf2025inefficiencies,bis2024tokenisationcpmi,fsb2024stabilitytokenisation}. For this reason, a taxonomy of RWA tokenization should not merely distinguish tokenized from non-tokenized assets. It should identify which lifecycle phases are transformed, which settlement model is adopted, which functions are implemented on-chain, and which governance and control mechanisms determine the resulting risk profile.

A complementary technical literature studies concrete architectures for tokenizing securities and other cashflow-bearing instruments. Li et al.~\cite{li2024tokenizingsecurities} propose a framework for tokenizing securities in decentralized finance ecosystems, combining fungible tokens with auxiliary smart contracts for rights management, redemption, and DeFi compatibility. Related work on dividend-bearing tokens shows that cashflow distribution rules are not merely implementation details, since accrual, distribution, and holder-identification mechanisms can affect incentives, secondary-market pricing, and legal compliance~\cite{zhitomirskiy2023dividendtoken}. These contributions motivate taxonomy dimensions related to accounting, servicing, redemption, and cashflow logic.

Legal and policy-oriented research further sharpens the classification problem. Lavayssi{\`e}re distinguishes between complete tokenization, where the token embodies legally enforceable rights, and incomplete tokenization, where the token acts primarily as a digital representation with limited standalone legal effect~\cite{lavayssiere2025legalstructures}. Policy and standard-setting reports emphasize legal uncertainty, fragmented standards, governance risk, identity and permissioning requirements, and the need for resilient settlement arrangements~\cite{bis2024tokenisationcpmi,fsb2024stabilitytokenisation,oecd2025impediments,iosco2025tokenization}. Together, these works show that technically similar token systems may differ substantially in custody, legal enforceability, investor protection, disclosure, compliance dependencies, and operational resilience.

Asset-class specific studies show that RWA tokenization cannot be fully captured by blockchain primitives alone. In real estate, tokenization is shaped by registry dependence, transfer restrictions, valuation practices, and jurisdiction-specific property law; systematic review evidence identifies regulatory fragmentation, organizational inertia, and technological risk as recurring barriers~\cite{eljaouhari2025realestateSLR}. In intellectual property, NFT-based patent tokenization raises distinct issues around provenance, valuation, licensing, and enforceability~\cite{bamakan2022patentnft,aksoy2025routledge}. These cases indicate that a useful taxonomy should separate cross-cutting tokenization dimensions, such as rights, custody, settlement, compliance, governance, and composability, from asset-class-specific constraints.

Several works explicitly attempt to organize the RWA and broader crypto-asset domain through taxonomies, archetypes, or classification frameworks. Earlier token taxonomies distinguish blockchain tokens according to broad functional, technical, and economic characteristics, including categories such as payment tokens, utility tokens, and asset or security tokens~\cite{oliveira2018token,schwiderowski2024crypto}. This classificatory perspective also helps avoid conflating RWA tokens with on-chain derivative instruments such as perpetual futures, which track a reference asset synthetically without creating a claim on, or custody relationship with, the underlying asset~\cite{pennella2026unifiedframeworkcomparativestudy}. These frameworks provide useful foundations for understanding token systems, but they remain too coarse for RWA tokenization because they do not capture the off-chain asset link, the custody model, the redemption mechanism, the enforceability of token-holder claims, or the role of legal wrappers.

More recent work focuses directly on tokenized real-world assets. Xia et al. provide a broad overview of RWA tokenization, including procedures, benefits, and open challenges~\cite{xia2025exploration}. Plepi and Schwendner propose tokenization archetypes and relate classification to potential efficiency and liquidity outcomes~\cite{plepi2024archetypes}. Aliyev proposes a taxonomy of RWA tokenisation methods that distinguishes between 1-for-1 ownership, 1-for-1 custodial structures, collateralised pools, and algorithmic or synthetic issuance models~\cite{aliyev2023taxonomy}. This classification is particularly useful for separating direct ownership claims from intermediated or collateral-based arrangements and for clarifying the role of backing, redemption, and trust assumptions.

Most closely related to our work, Luo et al.~\cite{luo2026sok} provide a systematization of knowledge of RWA tokenization, proposing a lifecycle-oriented taxonomy and a multi-layered architecture that spans legal custody, technical standards, cryptoeconomic valuation, oracle constraints, and jurisdictional interoperability. Their work offers a broad conceptual systematization of the RWA design space and highlights the tension between deterministic on-chain execution and off-chain legal, institutional, and informational dependencies. Our contribution differs by developing an attribute-based taxonomy designed for protocol-level classification and by applying it systematically to a sample of deployed RWA protocols. In doing so, we emphasize comparative classification, observable implementation choices, and cross-protocol design patterns across governance, asset structure, token properties, DLT, and economy.

Adjacent classification efforts also inform the present taxonomy. Zhang et al. develop a multidimensional crypto-asset taxonomy for investors and regulators, linking technical design, market structure, legal classification, minting, yield, and redemption mechanisms~\cite{zhang2026cryptoasset}. Their approach shows the value of asset-level classification for regulatory and market analysis, but their scope is broader than RWA tokenization and does not focus specifically on off-chain asset backing, legal wrappers, custody arrangements, and RWA-specific servicing logic. Similarly, work on blockchain-based investment intermediaries develops taxonomies and archetypes based on functionality, architecture, and security, showing how decentralized finance often produces new hybrid intermediary structures rather than eliminating intermediation altogether~\cite{zeiss2025intermediaries}. This perspective is relevant because many RWA systems operate as hybrid financial intermediaries that combine on-chain token representation with off-chain origination, custody, servicing, compliance, or reserve verification.

Compared with prior studies, our taxonomy operates at the system-architecture level. Rather than focusing narrowly on token types, asset classes, or issuance mechanisms, it captures how RWA systems integrate legal arrangements, asset structures, and DLT-based components. By situating this classification within the broader literature on blockchain-system taxonomies, the paper connects RWA-specific analysis to established debates on infrastructure and governance. This perspective enables RWA protocols to be compared as hybrid financial infrastructures, where on-chain code, off-chain institutions, legal claims, and market-access rules jointly shape the economic meaning of the token.

Overall, the literature reveals three main gaps that motivate a dedicated taxonomy for RWA tokenization. Existing studies tend to examine technical, legal, and policy dimensions separately; existing classifications often provide limited support for systematic comparison; and several taxonomies remain primarily descriptive rather than operational. Addressing these gaps requires a taxonomy that is both conceptually grounded and empirically applicable.

The novelty of our work therefore lies not in identifying tokenization as a phenomenon, but in providing a controlled system-level taxonomy that connects the complex legal, technical, operational, and market-access dimensions of RWA systems within a unified analytical framework.

\section{Methodology}
\label{sec:methodology}

This work follows the taxonomy-development method proposed by Nickerson et al.~\cite{nickerson2013method}, combining empirical-to-conceptual and conceptual-to-empirical iterations. The objects of interest were RWA tokenization protocols. In the initial empirical-to-conceptual step, a set of RWA systems was examined to identify recurring attributes relevant for comparing how off-chain assets are legally, economically, and technically represented on-chain. This first stage produced a broad list of candidate attributes covering legal structure, asset backing, tokenholder rights, token mechanics, on-chain implementation, fees, and access conditions.

The candidate attributes were then organized into five taxonomy components: \emph{Governance}, \emph{Asset}, \emph{Token Properties}, \emph{DLT}, and \emph{Economy}. \emph{Governance} captures issuer-related, legal, and control arrangements; \emph{Asset} captures the characteristics of the underlying off-chain asset and its backing structure; \emph{Token Properties} captures the design of the tokenized claim; \emph{DLT} captures the on-chain implementation; and \emph{Economy} captures fee structures and access conditions. The taxonomy was then iteratively applied to additional protocols, expanding the sample to twenty RWA systems. During these iterations, dimensions and permitted values were refined to improve consistency, comparability, and coverage across cases. The process was considered complete when the classification of additional systems did not require the introduction of substantially new dimensions, indicating sufficient coverage for the observed RWA protocol sample.

\section{Taxonomy}

The meta-characteristic of the proposed taxonomy is the set of design choices through which real-world assets are legally, economically, and technically represented on-chain. The taxonomy enables system-level comparison of RWA protocols by organizing twenty-three classification attributes into five components.

Several dimensions are implemented as controlled multi-label dimensions. This departs from the strictly mutually exclusive structure emphasized by Nickerson et al.~\cite{nickerson2013method}, but reflects the empirical nature of RWA systems: a single protocol may grant multiple entitlements, operate across several blockchains, charge different fee types, integrate with multiple DeFi protocols, or combine KYC, whitelisting, and qualified-investor restrictions. To preserve comparability, all multi-label assignments are restricted to predefined characteristics and are recorded only when explicitly supported by protocol documentation.

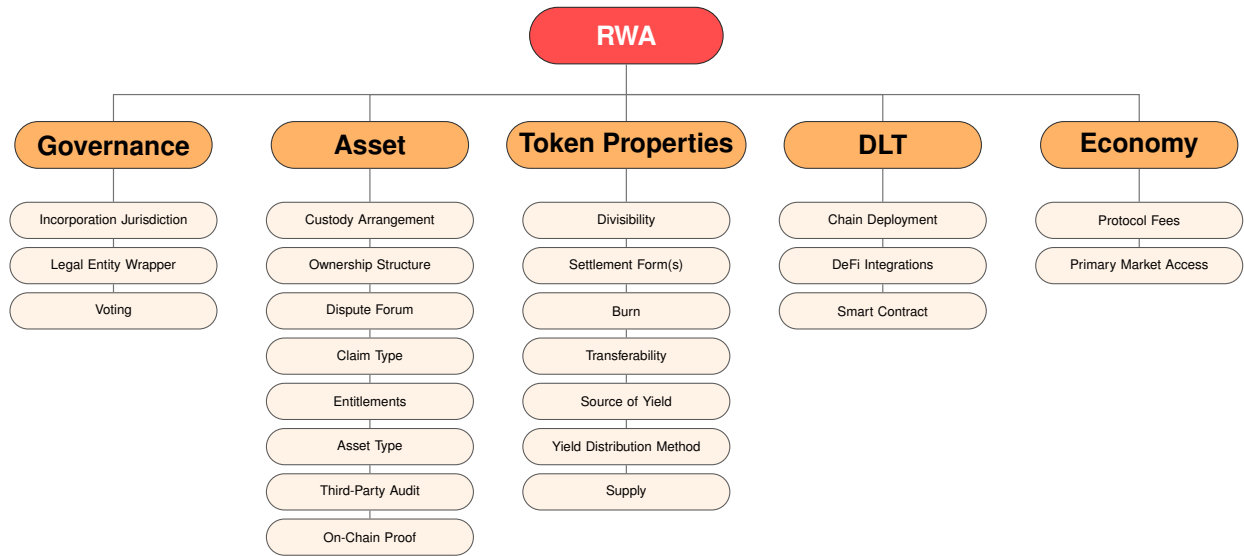
\begin{figure*}[t]
\centering
\resizebox{0.90\textwidth}{!}{%
\begin{tikzpicture}[
    font=\sffamily,
    taxonomy/.style={
        rounded rectangle,
        draw=black!85,
        fill=red!70,
        text=white,
        font=\sffamily\bfseries,
        minimum width=2.8cm,
        minimum height=0.75cm,
        align=center
    },
    component/.style={
        rounded rectangle,
        draw=black!80,
        fill=orange!60,
        font=\sffamily\bfseries,
        minimum width=2.85cm,
        minimum height=0.62cm,
        align=center
    },
    attribute/.style={
        rounded rectangle,
        draw=black!65,
        fill=orange!10,
        minimum width=2.85cm,
        minimum height=0.48cm,
        text width=2.70cm,
        align=center,
        font=\sffamily\tiny,
        inner sep=1.5pt
    },
    edge/.style={
        draw=black!55,
        line width=0.45pt
    }
]

\node[taxonomy] (tax) at (0,0) {RWA};

\node[component] (gov)   at (-6.8,-1.45) {Governance};
\node[component] (asset) at (-3.4,-1.45) {Asset};
\node[component] (token) at (0.0,-1.45) {Token Properties};
\node[component] (dlt)   at (3.4,-1.45) {DLT};
\node[component] (eco)   at (6.8,-1.45) {Economy};

\draw[edge] (tax.south) -- ++(0,-0.40) coordinate (hub);
\draw[edge] (hub) -| (gov.north);
\draw[edge] (hub) -| (asset.north);
\draw[edge] (hub) -- (token.north);
\draw[edge] (hub) -| (dlt.north);
\draw[edge] (hub) -| (eco.north);

\node[attribute] (gov1) at (-6.8,-2.45) {Incorporation Jurisdiction};
\node[attribute] (gov2) at (-6.8,-3.05) {Legal Entity Wrapper};
\node[attribute] (gov3) at (-6.8,-3.65) {Voting};

\draw[edge] (gov.south) -- (gov1.north);
\draw[edge] (gov1.south) -- (gov2.north);
\draw[edge] (gov2.south) -- (gov3.north);

\node[attribute] (a1) at (-3.4,-2.45) {Custody Arrangement};
\node[attribute] (a2) at (-3.4,-3.05) {Ownership Structure};
\node[attribute] (a3) at (-3.4,-3.65) {Dispute Forum};
\node[attribute] (a4) at (-3.4,-4.25) {Claim Type};
\node[attribute] (a5) at (-3.4,-4.85) {Entitlements};
\node[attribute] (a6) at (-3.4,-5.45) {Asset Type};
\node[attribute] (a7) at (-3.4,-6.05) {Third-Party Audit};
\node[attribute] (a8) at (-3.4,-6.65) {On-Chain Proof};

\draw[edge] (asset.south) -- (a1.north);
\draw[edge] (a1.south) -- (a2.north);
\draw[edge] (a2.south) -- (a3.north);
\draw[edge] (a3.south) -- (a4.north);
\draw[edge] (a4.south) -- (a5.north);
\draw[edge] (a5.south) -- (a6.north);
\draw[edge] (a6.south) -- (a7.north);
\draw[edge] (a7.south) -- (a8.north);

\node[attribute] (t1) at (0.0,-2.45) {Divisibility};
\node[attribute] (t2) at (0.0,-3.05) {Settlement Form(s)};
\node[attribute] (t3) at (0.0,-3.65) {Burn};
\node[attribute] (t4) at (0.0,-4.25) {Transferability};
\node[attribute] (t5) at (0.0,-4.85) {Source of Yield};
\node[attribute] (t6) at (0.0,-5.45) {Yield Distribution Method};
\node[attribute] (t7) at (0.0,-6.05) {Supply};

\draw[edge] (token.south) -- (t1.north);
\draw[edge] (t1.south) -- (t2.north);
\draw[edge] (t2.south) -- (t3.north);
\draw[edge] (t3.south) -- (t4.north);
\draw[edge] (t4.south) -- (t5.north);
\draw[edge] (t5.south) -- (t6.north);
\draw[edge] (t6.south) -- (t7.north);

\node[attribute] (d1) at (3.4,-2.45) {Chain Deployment};
\node[attribute] (d2) at (3.4,-3.05) {DeFi Integrations};
\node[attribute] (d3) at (3.4,-3.65) {Smart Contract};

\draw[edge] (dlt.south) -- (d1.north);
\draw[edge] (d1.south) -- (d2.north);
\draw[edge] (d2.south) -- (d3.north);

\node[attribute] (e1) at (6.8,-2.45) {Protocol Fees};
\node[attribute] (e2) at (6.8,-3.05) {Primary Market Access};

\draw[edge] (eco.south) -- (e1.north);
\draw[edge] (e1.south) -- (e2.north);

\end{tikzpicture}%
}
\vspace{-0.5em}
\caption{Overview of the proposed RWA tokenization taxonomy. Lines indicate conceptual association only; attributes listed under the same component are co-equal and their order does not imply hierarchy, sequence, or dependency.}
\label{fig:rwa-taxonomy-overview}
\vspace{-1em}
\end{figure*}

Figure~\ref{fig:rwa-taxonomy-overview} presents the taxonomy as a tree-like overview, grouping the twenty-three dimensions under the five components. The full taxonomy matrix, including definitions and permitted characteristics for each dimension, is reported in Appendix~\ref{app:rwa-taxonomy-matrix}.

The following subsections describe each component in greater detail.

\subsection{Governance}

The \emph{Governance} component captures the legal and institutional setting of the RWA protocol. Incorporation jurisdiction identifies the country or territory in which the issuing entity is formally registered, as this determines the applicable corporate and regulatory environment. Legal entity wrapper describes the legal form through which the tokenized asset is issued, distinguishing direct issuers from bankruptcy-remote vehicles, trusts, Special Purpose Vehicles (SPVs), or other structures. Voting captures whether tokenholders or investors are granted any governance rights, and, where applicable, the form of such voting mechanisms.

\subsection{Asset}

The \emph{Asset} component describes the off-chain asset and the legal or economic relationship between the tokenholder and that asset. Custody arrangement identifies how the underlying asset is safeguarded, for example through a regulated custodian, trust arrangement, issuer self-custody, or physical vault. Ownership structure captures who legally owns the underlying asset and whether the tokenholder holds direct title, beneficial ownership, an SPV-mediated interest, or no legal title. Dispute forum records the governing jurisdiction or dispute-resolution venue applicable to claims. Claim type classifies the nature of the tokenholder’s claim, such as a direct claim, pro-rata claim, synthetic exposure, or creditor claim. Entitlements identifies the actionable rights attached to the token, including redemption rights, income distributions, interest entitlement, governance rights, or no additional rights beyond ownership. Asset type classifies the underlying real-world asset, such as fixed income, equities, commodities, real estate, carbon credits, or collectibles. Finally, verification mechanisms capture whether the backing of the token is externally verified through third-party audits, attestations, proof-of-reserve systems, or on-chain oracle feeds.

\subsection{Token Properties}

The Token Properties component captures how the RWA claim is technically represented and operationalized as a blockchain token. Divisibility records whether the token can be held fractionally or only in whole units. Settlement Form(s) captures the consideration received upon redemption, such as cash, the underlying asset, another token, or a combination. Burn records whether the token is burned or otherwise cancelled upon redemption. Transferability identifies restrictions on secondary-market transfers, including unrestricted transfers, whitelist restrictions, lock-up periods, issuer freeze capabilities, or non-transferability. Source of yield identifies the economic source of returns, such as interest income, coupon payments, dividends, rental income, lending interest, or price appreciation only. Yield distribution method describes how returns are delivered to holders, for example through accrual, rebasing, distribution, streaming, buyback-and-burn, or yield stripping. Supply captures whether the token supply is capped or uncapped.

\subsection{Distributed Ledger Technology}

The DLT component captures the blockchain infrastructure and on-chain implementation of the RWA system. Chain deployment records the blockchain networks on which the token is officially issued or supported by the issuer. DeFi integrations identify whether and how the token is used in decentralized finance, including as collateral, in liquidity pools, structured products, vaults, staking systems, or other integrations. The smart contract dimension captures which functions of the RWA system are implemented or automated on-chain, including minting, redemption, transfer restrictions, whitelist enforcement, oracle pricing, proof-of-reserve reporting, yield distribution, or other protocol functions. This component distinguishes protocols where blockchain is primarily used as a recordkeeping and distribution layer from those where core operational processes and market integrations are executed through on-chain code.

\subsection{Economy}

The Economy component captures the user-facing economic and access conditions of the RWA system. The Protocol Fees dimension records the types of fees charged to users, such as minting fees, redemption fees, custody fees, management fees, transfer taxes, or yield spreads. The Primary Market Access dimension identifies who can acquire tokens directly from the issuer and under what eligibility conditions, including permissionless access, KYC-gated access, qualified-investor restrictions, institutional-only access, or whitelist requirements.

The complete list of permitted characteristics for each dimension is reported in the taxonomy matrix in the Appendix. During classification, protocols are mapped only to these predefined characteristics. Where documentation does not explicitly verify a characteristic, the value is recorded as \textit{N/A}; where a protocol exhibits a relevant characteristic not covered by the taxonomy, it is recorded as \textit{Other} and documented separately.

\section{Classification}

The taxonomy was applied to the twenty largest tokenized real-world asset protocols listed on DeFiLlama's RWA dashboard by market capitalization in February 2026. The use of DeFiLlama provides an external selection criterion and reduces arbitrariness in the construction of the sample. The resulting dataset should therefore be understood as a market-capitalization-based snapshot of leading RWA tokenization protocols at the time of data collection, rather than as a random sample of the entire RWA universe.
For each protocol, classification was based on publicly available documentation, including official protocol documentation, legal documents, product pages, smart-contract references, audit or attestation reports, and, where relevant, on-chain information. Each classification entry is linked to a source citation in the protocol repository.
\footnote{https://github.com/GiorgioVella/rwa-tokenization-taxonomy}.

The classification results are reported component by component, making it possible to compare protocols horizontally within a single component while preserving the internal structure of the taxonomy. Tables~\ref{tab:rwa-classification-governance}--\ref{tab:rwa-classification-economy} therefore report the classifications for the \emph{Governance}, \emph{Asset}, \emph{Token Properties}, \emph{DLT}, and \emph{Economy} components.

Table~\ref{tab:rwa-classification-governance} reports the \emph{Governance} component. The protocols are predominantly organized through conventional legal entities, including direct issuers, trusts, SPVs, limited partnerships, and limited liability companies. Jurisdictions are concentrated in established financial or fund-structuring venues such as the United States, the British Virgin Islands, France, Bermuda, Panama, Germany, and El Salvador. Voting rights are generally absent or not publicly documented at the token level. Where voting appears, it is typically attached to shareholder or fund governance rather than to an on-chain governance mechanism. This supports the interpretation that leading RWA protocols remain legally mediated and issuer-administered rather than DAO-governed.

\begin{table*}[!t]
\centering
\scriptsize
\setlength{\tabcolsep}{2pt}
\renewcommand{\arraystretch}{1.08}
\caption{Protocol classification according to the Governance component.}
\label{tab:rwa-classification-governance}
\begin{adjustbox}{max width=\textwidth}
\begin{tabular}{@{}r p{3.1cm} p{1.6cm} p{2.4cm} p{3.0cm}@{}}
\toprule
\textbf{ID} & \textbf{protocol} & \textbf{Jurisdiction} & \textbf{Legal Entity} & \textbf{Voting} \\
\midrule
1 & Tether Gold & El Salvador & Direct issuer & N/A \\
2 & BUIDL & BVI & Direct issuer & N/A \\
3 & PAX Gold & USA & Direct issuer & N/A \\
4 & US Yield Coin & Bermuda & Direct issuer & N/A \\
5 & Superstate USTB & USA & Trust & N/A \\
6 & WisdomTree Treasury MMF & USA & Trust & Shareholder voting \\
7 & Anemoy AAA CLO Fund & BVI & SPV & N/A \\
8 & Anemoy Treasury Fund & BVI & SPV & N/A \\
9 & Ondo OUSG & USA & Limited partnership & N/A \\
10 & BUIDL-I & BVI & Direct issuer & N/A \\
11 & Hastra PRIME & BVI & LLC & N/A \\
12 & Franklin BENJI & USA & Trust & Shareholder voting \\
13 & Superstate USCC & USA & Trust & N/A \\
14 & Exodus Class B & USA & Direct issuer & N/A \\
15 & Theo thBILL & Panama & Direct issuer & N/A \\
16 & Fidelity FDIT & USA & Trust & N/A \\
17 & Spiko EUTBL & France & SPV & N/A \\
18 & Spiko USTBL & France & SPV & N/A \\
19 & Mantle MI4 & BVI & SPV & N/A \\
20 & mF-ONE & Germany & Direct issuer & N/A \\
\bottomrule
\end{tabular}
\end{adjustbox}
\end{table*}

Table~\ref{tab:rwa-classification-asset} reports the \emph{Asset} component. The sample is strongly concentrated in financial assets, especially fixed-income instruments, money-market funds, Treasury-bill strategies, credit products, and other yield-bearing fund structures. Only a small number of protocols tokenize commodities or equity-like claims. Most protocols rely on institutional custody, vault storage, SPV-held assets, trusts, or other off-chain legal arrangements. Tokenholder claims are commonly structured as pro-rata claims on pooled assets, usually combined with redemption rights and income or interest entitlements. This shows that the token does not replace the legal ownership structure; rather, it represents an on-chain interface to an off-chain financial and legal claim.

\begin{table*}[!t]
\centering
\tiny
\setlength{\tabcolsep}{1.4pt}
\renewcommand{\arraystretch}{1.05}
\caption{Protocol classification according to the Asset component.}
\label{tab:rwa-classification-asset}
\begin{adjustbox}{max width=\textwidth}
\begin{tabular}{@{}r p{2.0cm} p{1.6cm} p{1.8cm} p{1.6cm} p{1.6cm} p{1.6cm} p{1.5cm} p{1.4cm} p{1.3cm}@{}}
\toprule
\textbf{ID} & \textbf{protocol} & \textbf{Custody} & \textbf{Ownership} & \textbf{Dispute Forum} & \textbf{Claim Type} & \textbf{Entitlements} & \textbf{Asset Type} & \textbf{Audit} & \textbf{On-chain Proof} \\
\midrule
1 & Tether Gold & Vault storage & Beneficial ownership & BVI arbitration & Direct claim & Redemption & Commodity & Yes & Other \\
2 & BUIDL & Institutional custodian & Beneficial ownership & Delaware / New York & Pro-rata claim & Redemption; income & Fixed income & Yes & N/A \\
3 & PAX Gold & Vault storage & Beneficial ownership & New York arbitration & Pro-rata claim & Redemption & Commodity & Yes & Other \\
4 & US Yield Coin & Institutional custodian & Beneficial ownership & N/A & Pro-rata claim & Redemption; interest & Fixed income & Yes & Oracle feed \\
5 & Superstate USTB & Institutional custodian & SPV held & N/A & Pro-rata claim & Redemption; interest & Fixed income & Yes & Oracle feed \\
6 & WisdomTree Treasury MMF & Institutional custodian & SPV held & New York arbitration & Pro-rata claim & Redemption; income; voting & Fixed income & Yes & N/A \\
7 & Anemoy AAA CLO Fund & Institutional custodian & SPV held & N/A & Pro-rata claim & Redemption; interest & Fixed income & Yes & Other \\
8 & Anemoy Treasury Fund & Institutional custodian & Direct title & N/A & Pro-rata claim & Redemption; interest & Fixed income & Yes & Other \\
9 & Ondo OUSG & Institutional custodian & LP interest & N/A & Pro-rata claim & Redemption; interest & Fixed income & Yes & Oracle feed \\
10 & BUIDL-I & Institutional custodian & SPV held & N/A & Pro-rata claim & Redemption; income; interest & Fixed income & Yes & N/A \\
11 & Hastra PRIME & Trust arrangement & Beneficial ownership & N/A & Pro-rata claim & Redemption; interest & Private credit & N/A & Other \\
12 & Franklin BENJI & Institutional custodian & Direct title & N/A & Pro-rata claim & Redemption; income; voting & Fixed income & Yes & N/A \\
13 & Superstate USCC & Institutional custodian & SPV held & N/A & Pro-rata claim & Redemption; interest & Crypto carry & Yes & Oracle feed \\
14 & Exodus Class B & N/A & Direct title & USA / Texas & Direct claim & Income; voting & Equity & N/A & N/A \\
15 & Theo thBILL & Institutional custodian & N/A & Panama arbitration & Synthetic exposure & Redemption; interest & Fixed income & N/A & Oracle feed \\
16 & Fidelity FDIT & Institutional custodian & SPV held & N/A & Pro-rata claim & Redemption; income; voting & Fixed income & Yes & N/A \\
17 & Spiko EUTBL & Institutional custodian & SPV held & France & Pro-rata claim & Redemption; interest; voting & Fixed income & Yes & Oracle feed \\
18 & Spiko USTBL & Institutional custodian & SPV held & France & Pro-rata claim & Redemption; interest; voting & Fixed income & Yes & Oracle feed \\
19 & Mantle MI4 & Self-custodied & SPV held & N/A & Pro-rata claim & Redemption & Crypto index & Yes & N/A \\
20 & mF-ONE & N/A & N/A & N/A & Debt claim & Redemption & Other & N/A & Oracle feed \\
\bottomrule
\end{tabular}
\end{adjustbox}
\end{table*}

The \emph{Token Properties} component, reported in Table~\ref{tab:rwa-classification-token-properties}, shows that most tokens are fractionalized and transferable only under some form of restriction, most commonly whitelist-based transferability or issuer-controlled transfer rules. Redemption is usually available, although the settlement form differs across protocols, including cash, stablecoin or token settlement, physical delivery, and mixed settlement mechanisms. Yield-bearing protocols most often rely on interest income and deliver returns through accrual, distribution, or rebasing mechanisms. Several entries remain classified as \textit{N/A}, especially for burn mechanics and supply constraints, indicating that public documentation does not always describe the token lifecycle with sufficient precision.

\begin{table*}[!t]
\centering
\tiny
\setlength{\tabcolsep}{1.5pt}
\renewcommand{\arraystretch}{1.05}
\caption{Protocol classification according to the Token Properties component.}
\label{tab:rwa-classification-token-properties}
\begin{adjustbox}{max width=\textwidth}
\begin{tabular}{@{}r p{2.0cm} p{1.4cm} p{1.4cm} p{1.1cm} p{1.8cm} p{1.5cm} p{1.4cm} p{1.5cm}@{}}
\toprule
\textbf{ID} & \textbf{protocol} & \textbf{Divisibility} & \textbf{Settlement} & \textbf{Burn} & \textbf{Transferability} & \textbf{Yield Source} & \textbf{Yield Method} & \textbf{Supply} \\
\midrule
1 & Tether Gold & Fractionalized & Combination & Yes & Issuer freeze & N/A & N/A & Uncapped \\
2 & BUIDL & Fractionalized & Combination & Yes & Whitelist & Interest & Rebase & Dynamic \\
3 & PAX Gold & Fractionalized & Combination & Yes & Issuer freeze & Price appreciation & N/A & Dynamic \\
4 & US Yield Coin & Fractionalized & In-kind & Yes & Whitelist & Interest & Accrual & N/A \\
5 & Superstate USTB & Fractionalized & Combination & Yes & Whitelist & Interest & Accrual & Uncapped \\
6 & WisdomTree Treasury MMF & Fractionalized & Combination & Yes & Whitelist & Interest & Distribution & Uncapped \\
7 & Anemoy AAA CLO Fund & Fractionalized & In-kind & Yes & Conditional & Interest & Accrual & Uncapped \\
8 & Anemoy Treasury Fund & Fractionalized & Combination & Yes & Whitelist & Interest & Accrual & Uncapped \\
9 & Ondo OUSG & Fractionalized & In-kind & N/A & Whitelist & Interest & Accrual & N/A \\
10 & BUIDL-I & Fractionalized & N/A & N/A & Whitelist & Interest & Distribution & Uncapped \\
11 & Hastra PRIME & Fractionalized & In-kind & N/A & Unrestricted & Lending interest & Accrual & N/A \\
12 & Franklin BENJI & Fractionalized & Cash & N/A & Whitelist & Interest & Rebase & N/A \\
13 & Superstate USCC & Fractionalized & Combination & Yes & Whitelist & Basis / staking & Accrual & N/A \\
14 & Exodus Class B & N/A & N/A & N/A & Whitelist & Price appreciation & N/A & Capped \\
15 & Theo thBILL & Fractionalized & In-kind & N/A & Unrestricted & Interest & Accrual & N/A \\
16 & Fidelity FDIT & Fractionalized & Cash & N/A & Whitelist & Interest & Distribution & N/A \\
17 & Spiko EUTBL & Fractionalized & Combination & Yes & Whitelist & Interest & Accrual & Uncapped \\
18 & Spiko USTBL & Fractionalized & Combination & Yes & Whitelist & Interest & Accrual & Uncapped \\
19 & Mantle MI4 & N/A & N/A & N/A & Whitelist & Staking rewards & Accrual & Uncapped \\
20 & mF-ONE & Fractionalized & N/A & N/A & Whitelist & Other & Accrual & N/A \\
\bottomrule
\end{tabular}
\end{adjustbox}
\end{table*}

Table~\ref{tab:rwa-classification-dlt} reports the \emph{DLT} component. The observed protocols use DLT primarily for token issuance, transfer permissioning, recordkeeping, redemption or burn workflows, oracle-based pricing, and cross-chain deployment. Ethereum remains the most recurrent deployment environment, but several protocols support additional networks such as Solana, Polygon, Arbitrum, Base, Avalanche, Stellar, and other chains. DeFi integration is heterogeneous: some tokens are used as collateral in lending markets or exchange venues, while for others no DeFi integration could be verified from public documentation.

\begin{table*}[!t]
\centering
\tiny
\setlength{\tabcolsep}{1.5pt}
\renewcommand{\arraystretch}{1.05}
\caption{Protocol classification according to the Distributed Ledger Technology component.}
\label{tab:rwa-classification-dlt}
\begin{adjustbox}{max width=\textwidth}
\begin{tabular}{@{}r p{2.1cm} p{4.6cm} p{2.8cm} p{3.2cm}@{}}
\toprule
\textbf{ID} & \textbf{protocol} & \textbf{Smart Contract Functions} & \textbf{Chain Deployment} & \textbf{DeFi Integrations} \\
\midrule
1 & Tether Gold & Transfer; mint/burn; freeze; revoke & Ethereum & Collateral \\
2 & BUIDL & Tokenized shares; whitelist; mint/burn; freeze; USDC swap & Ethereum; Aptos; Arbitrum; Avalanche; Optimism; Polygon; Solana; BNB & Collateral; UniswapX \\
3 & PAX Gold & Transfer; freeze; upgrade; allocation lookup & Ethereum; Solana & DEX liquidity \\
4 & US Yield Coin & Mint; redeem; permissioning; oracle pricing & Ethereum; BSC; Solana & Collateral \\
5 & Superstate USTB & Mint; redeem; permissioning; bridge; oracle pricing & Ethereum; Solana; Plume & Aave \\
6 & WisdomTree Treasury MMF & Issuance; recordkeeping; permissioning; P2P transfer & Ethereum; Stellar; Avalanche; Arbitrum; Base; Optimism; Plume & N/A \\
7 & Anemoy AAA CLO Fund & Mint; redeem; permissioning; NAV oracle; multichain issuance & Ethereum; Base; Avalanche; BSC; Solana & Aave Horizon \\
8 & Anemoy Treasury Fund & Mint; redeem; permissioning; NAV oracle; multichain issuance & Ethereum; Base; Arbitrum; Celo & Aave Horizon \\
9 & Ondo OUSG & Mint; redeem; permissioning; price oracle; multichain issuance & Ethereum; Polygon; Solana; XRP Ledger & Flux Finance \\
10 & BUIDL-I & ERC-20 proxy; permissioning; USDC off-ramp & Ethereum & Collateral; liquidity \\
11 & Hastra PRIME & Staking; unstaking; redemption; transfer & Solana & Lending collateral \\
12 & Franklin BENJI & Issuance; recordkeeping; permissioning; P2P transfer; yield distribution & Stellar; Polygon; Arbitrum; Avalanche; Aptos; Ethereum; Base; Solana; BNB & Binance collateral \\
13 & Superstate USCC & Mint; redeem; permissioning; bridge; oracle pricing & Ethereum; Solana; Plume & Aave; Kamino; Morpho \\
14 & Exodus Class B & Issuance; maintenance; permissioning; P2P transfer & Arbitrum & N/A \\
15 & Theo thBILL & Mint; redeem; basket controls; oracle pricing; bridge & Ethereum; Arbitrum; Base; HyperEVM & Lending; DEX; derivatives \\
16 & Fidelity FDIT & Share recordkeeping; issuance; redemption; permissioning & Ethereum & N/A \\
17 & Spiko EUTBL & Share register; permissioning; NAV oracle; redemption/burn; bridge & Ethereum; Polygon; Arbitrum; Starknet; Base; Etherlink; Stellar & Morpho; SG Forge \\
18 & Spiko USTBL & Share register; permissioning; NAV oracle; redemption/burn; bridge & Ethereum; Polygon; Arbitrum; Starknet; Base; Etherlink; Stellar & Morpho; SG Forge \\
19 & Mantle MI4 & Fund tokenization; permissioning; transfer; collateral use & Mantle Network & Exchange collateral \\
20 & mF-ONE & ERC-20 token; issuance vault; redemption vault; NAV oracle & Ethereum & Liquidity pool; Morpho \\
\bottomrule
\end{tabular}
\end{adjustbox}
\end{table*}

Finally, the \emph{Economy} component, reported in Table~\ref{tab:rwa-classification-economy}, shows that primary-market access is mostly permissioned. KYC, whitelisting, qualified-investor requirements, and institutional access restrictions are common even when tokens are issued on public blockchains. Fee structures also vary, including minting and redemption fees, management fees, custody fees, yield spreads, DLT transaction fees, and operating expenses. The classification therefore reveals a hybrid access model: blockchain infrastructure may support transferability and composability, but primary issuance and redemption remain shaped by legal, regulatory, and commercial constraints.

\begin{table*}[!t]
\centering
\scriptsize
\setlength{\tabcolsep}{2pt}
\renewcommand{\arraystretch}{1.08}
\caption{Protocol classification according to the Economy component.}
\label{tab:rwa-classification-economy}
\begin{adjustbox}{max width=\textwidth}
\begin{tabular}{@{}r p{3.2cm} p{4.8cm} p{4.5cm}@{}}
\toprule
\textbf{ID} & \textbf{protocol} & \textbf{Protocol Fees} & \textbf{Primary Market Access} \\
\midrule
1 & Tether Gold & Minting; redemption & KYC gated \\
2 & BUIDL & Yield spread & Qualified investors only \\
3 & PAX Gold & Minting; redemption & KYC gated \\
4 & US Yield Coin & Minting; redemption; yield spread & Institutional; KYC; whitelist \\
5 & Superstate USTB & Custody fee & Qualified investors; KYC; whitelist \\
6 & WisdomTree Treasury MMF & Expense ratio; gas fees & KYC; whitelist \\
7 & Anemoy AAA CLO Fund & Management fee & Qualified investors only \\
8 & Anemoy Treasury Fund & Management fee & Qualified investors; whitelist \\
9 & Ondo OUSG & Management fee & Qualified investors; KYC; whitelist \\
10 & BUIDL-I & Flat fee; NAV-based fee & Qualified / pre-approved investors \\
11 & Hastra PRIME & N/A & KYC gated \\
12 & Franklin BENJI & Management fee & KYC; whitelist \\
13 & Superstate USCC & Custody; administration; audit; management & Qualified investors; KYC; whitelist \\
14 & Exodus Class B & No disclosed protocol fees & KYC; whitelist \\
15 & Theo thBILL & N/A & Institutional; KYC \\
16 & Fidelity FDIT & Management fee & Institutional; KYC \\
17 & Spiko EUTBL & Management; DLT transaction; operating; remediation & KYC; whitelist \\
18 & Spiko USTBL & Management; DLT transaction; operating; remediation & KYC; whitelist \\
19 & Mantle MI4 & Management fee & Qualified investors only \\
20 & mF-ONE & Minting; redemption & Qualified investors; KYC; whitelist \\
\bottomrule
\end{tabular}
\end{adjustbox}
\end{table*}

Taken together, the component-level tables show that leading RWA tokenization protocols share several common design patterns. They are mostly financial-asset protocols, rely heavily on off-chain legal and custodial infrastructure, and use public blockchains selectively for representation, transfer, settlement, and composability. At the same time, the sample displays meaningful variation in legal wrapper, ownership structure, redemption design, transfer restrictions, yield distribution, chain deployment, DeFi integration, and fee design.

The classification also reveals persistent documentation gaps. Several dimensions, including voting, dispute forum, burn mechanics, supply constraints, and verification mechanisms, are frequently difficult to verify from public documentation. This is analytically relevant: the taxonomy does not only classify observable design choices, but also highlights areas where RWA protocols provide insufficient public information. Overall, the classification supports the view that RWA tokenization is not a single uniform model, but a set of hybrid legal-financial-technical configurations.

\section{Conclusion}

This paper proposed a protocol-level taxonomy for real-world asset tokenization systems that jointly captures legal, technical, and economic design choices. By organizing twenty-three dimensions across governance, asset structure, token properties, DLT, and economy, the taxonomy provides a structured basis for comparing how RWA protocols represent off-chain assets, define tokenholder claims, implement custody and verification mechanisms, restrict transfers, distribute yield, and integrate with DeFi protocols. Its application to twenty leading protocols shows that RWA tokenization is best understood as a hybrid form of financial infrastructure: public blockchains and smart contracts support representation, transfer control, redemption workflows, pricing, and composability, while core guarantees remain anchored in legal wrappers, institutional custody, compliance processes, and reserve verification. These findings highlight the need to evaluate RWA systems across both on-chain and off-chain dimensions rather than treating tokenization as a purely technical transformation. Future work can extend the taxonomy to a larger and more diverse sample, examine changes over time, and assess how specific design choices affect liquidity, risk, interoperability, decentralization, and regulatory compliance.

\section*{Acknowledgment}

The contribution of Luca Pennella was funded in part by the PayPal-FNR PEARL Chair in Digital Financial Services, FNR grant reference 13342933/Gilbert Fridgen, and by the FutureFinTech National Centre of Excellence in Research and Innovation, FNR grant reference 16570468, with the support of Luxembourg’s Ministry of Finance. For the purpose of open access, and in fulfilment of the obligations arising from the grant agreements, the author has applied a Creative Commons Attribution 4.0 International (CC BY 4.0) license to any Author Accepted Manuscript version arising from this submission.

\section*{Data Availability}
The datasets and sources used for the experiments presented in this article are available online at \hyperlink{https://github.com/GiorgioVella/rwa-tokenization-taxonomy}{https://github.com/GiorgioVella/rwa-tokenization-taxonomy}

\bibliographystyle{IEEEtranN}

\bibliography{reference}

@misc{defillama_rwa,
  author       = {{DefiLlama}},
  title        = {{Real World Assets}},
  year         = {2026},
  url          = {https://defillama.com/rwa},
  note         = {Accessed: 2026-05-11}
}

@misc{xia2025exploration,
  author = {Xia, Wenchao and Zhang, Yike and Cheng, Yufei},
  title = {Exploration on Real World Assets and Tokenization},
  year = {2025},
  month = {may},
  doi = {10.48550/arXiv.2503.01111},
  url = {https://arxiv.org/abs/2503.01111},
  eprint = {2503.01111},
  archivePrefix = {arXiv},
  primaryClass = {cs.CY}
}

@inproceedings{li2024tokenizingsecurities,
  author = {Li, Zhilin and Li, Zijian and Wu, Kaixiang and Zhang, Danyu and Zhang, Ke and Liu, Donghan and Guo, Ziqi and Zhou, Chao and Guan, Hui and Niu, Jiyuan},
  title = {On Tokenizing Securities in Contemporary Decentralized Finance Ecosystems},
  booktitle = {2024 IEEE International Conference on Blockchain Research \& Applications for Innovative Networks and Services (BRAINS)},
  year = {2024},
  pages = {1--9},
  doi = {10.1109/BRAINS63024.2024.10732268},
  url = {https://doi.org/10.1109/BRAINS63024.2024.10732268},
  publisher = {IEEE}
}

@techreport{imf2025inefficiencies,
  author = {Agur, Itai and Villegas-Bauer, Germ{\'a}n and Mancini-Griffoli, Tommaso and {Martinez Peria}, Maria Soledad},
  title = {Tokenization and Financial Market Inefficiencies},
  institution = {International Monetary Fund},
  type = {Fintech Notes},
  number = {2025/001},
  year = {2025},
  month = {jan},
  pages = {33},
  doi = {10.5089/9798400298905.063}
}

@misc{plepi2024archetypes,
  author = {Plepi, Anisa and Schwendner, Peter},
  title = {A Method for Uncovering Tokenisation Archetypes and their Effects: Thus Spoke Switzerland},
  year = {2024},
  month = {nov},
  howpublished = {SSRN Working Paper},
  note = {SSRN eLibrary, accessed 2026-01-21}
}

@techreport{bis2024tokenisationcpmi,
  author = {Committee on Payments and Market Infrastructures and Bank for International Settlements},
  title = {Tokenisation in the Context of Money and Other Assets: Concepts and Implications for Central Banks},
  institution = {Bank for International Settlements},
  year = {2024},
  month = {oct},
  url = {https://www.bis.org/cpmi/publ/d225.htm}
}

@misc{fsb2024stabilitytokenisation,
  author = {Financial Stability Board},
  title = {The Financial Stability Implications of Tokenisation},
  year = {2024},
  month = {oct},
  howpublished = {FSB statement / web publication},
  url = {https://www.fsb.org/2024/10/the-financial-stability-implications-of-tokenisation/}
}

@techreport{oecd2025impediments,
  author = {OECD},
  title = {Tokenisation of Assets and Distributed Ledger Technologies in Financial Markets: Potential Impediments to Market Development and Policy Implications},
  institution = {OECD Publishing},
  type = {OECD Business and Finance Policy Papers},
  number = {75},
  address = {Paris},
  year = {2025},
  month = {jan},
  pages = {27},
  doi = {10.1787/40e7f217-en}
}

@techreport{iosco2025tokenization,
  author = {International Organization of Securities Commissions},
  title = {Tokenization of Financial Assets},
  institution = {International Organization of Securities Commissions},
  type = {Final Report},
  number = {FR/17/2025},
  year = {2025},
  month = {nov},
  url = {https://www.iosco.org/library/pubdocs/pdf/IOSCOPD809.pdf}
}

@article{lavayssiere2025legalstructures,
  author = {Lavayssi{\`e}re, Xavier},
  title = {Legal Structures of Tokenised Assets},
  journal = {European Journal of Risk Regulation},
  year = {2024},
  pages = {1--13},
  doi = {10.1017/err.2024.88}
}

@article{zhitomirskiy2023dividendtoken,
  author = {Zhitomirskiy, Vladislav and Franken, Hans and Schmidt, Niels and Schiller, Tobias and Hanke, Michael},
  title = {Tokenizing Assets with Dividend Payouts---A Legally Compliant and Flexible Design},
  journal = {Digital Finance},
  year = {2023},
  volume = {5},
  number = {3--4},
  pages = {1--18},
  doi = {10.1007/s42521-023-00094-w},
  url = {https://link.springer.com/article/10.1007/s42521-023-00094-w}
}

@article{eljaouhari2025realestateSLR,
  author = {{El Jaouhari}, Asmae and {\v{S}}e{\v{s}}plaukis, Tomas and L{\'a}zaro-Touza, Laura and Dagilyt{\.{e}}, Aurelija and Sk{\"a}rdi, Maria},
  title = {Tokenization and the Future of Property Investment: A New Paradigm for Real Estate},
  journal = {International Journal of Strategic Property Management},
  year = {2025},
  volume = {29},
  number = {4},
  pages = {297--315},
  doi = {10.3846/ijspm.2025.24814},
  url = {https://journals.vilniustech.lt/index.php/IJSPM/article/view/24814}
}

@article{bamakan2022patentnft,
  author = {Bamakan, Seyed Mojtaba Hosseini and Khajavi, Seyed H. and Manshadi, Seyed D. and Motlagh, Mahdi},
  title = {Tokenizing Patents and Intellectual Property Assets: Using Blockchain-Based Systems and Non-Fungible Tokens to Promote Financialization},
  journal = {Scientific Reports},
  year = {2022},
  volume = {12},
  pages = {2178},
  doi = {10.1038/s41598-022-05920-6},
  url = {https://www.nature.com/articles/s41598-022-05920-6}
}

@article{luo2026sok,
  title={SoK of RWA Tokenization: A Systematization of Concepts, Architectures, and Legal Interoperability},
  author={Luo, Junliang and Xiong, Xihan and Li, Zonglun and Kang, Hong and Liu, Xue and Knottenbelt, William J and Tinn, Katrin},
  journal={arXiv preprint arXiv:2604.06608},
  year={2026}
}

@article{aliyev2023taxonomy,
  title={A taxonomy of tokenisation methods for real-world assets},
  author={Aliyev, Nihad},
  journal={Available at SSRN 5214610},
  year={2023}
}

@article{zhang2026cryptoasset,
  title={Crypto-asset Taxonomy for Investors and Regulators},
  author={Zhang, Xiao and Iba{\~n}ez, Juan Ignacio and Xu, Jiahua},
  journal={arXiv preprint arXiv:2602.05098},
  year={2026}
}

@article{zeiss2025intermediaries,
  title={Exploring trust in decentralized finance intermediaries: a taxonomy and archetypes for guiding blockchain-based investment decisions on the web},
  author={Zei{\ss}, Christian and Straub, Lisa and Greiner, Maximilian and Neis, Marcel and Neis, Nicolas and Winkelmann, Axel and Lechner, Ulrike},
  journal={Internet Research},
  volume={35},
  number={7},
  pages={133--152},
  year={2025},
  publisher={Emerald Publishing Limited}
}

@inproceedings{oliveira2018token,
  title        = {To Token or not to Token: Tools for Understanding Blockchain Tokens},
  author       = {Oliveira, Luis and Zavolokina, Liudmila and Bauer, Ingrid and Schwabe, Gerhard},
  booktitle    = {Proceedings of the 39th International Conference on Information Systems},
  year         = {2018}
}

@article{schwiderowski2024crypto,
  title={Crypto tokens and token systems},
  author={Schwiderowski, Jan and Pedersen, Asger Balle and Beck, Roman},
  journal={Information Systems Frontiers},
  volume={26},
  number={1},
  pages={319--332},
  year={2024},
  publisher={Springer}
}

@article{nickerson2013method,
  title={A method for taxonomy development and its application in information systems},
  author={Nickerson, Robert C and Varshney, Upkar and Muntermann, Jan},
  journal={European journal of information systems},
  volume={22},
  number={3},
  pages={336--359},
  year={2013},
  publisher={Taylor \& Francis}
}

@article{pennella2026unifiedframeworkcomparativestudy,
  title={A Unified Framework and Comparative Study of Decentralized Finance Derivatives Protocols},
  author={Pennella, Luca and Saggese, Pietro and Pinelli, Fabio and Galletta, Letterio},
  journal={arXiv preprint arXiv:2512.19113},
  year={2025}
}

@book{aksoy2025routledge,
  title={Routledge Handbook of NFT Law},
  author={Aksoy, P{\i}nar {\c{C}}a{\u{g}}layan and Dimatteo, Larry and Hufnagel, Saskia},
  year={2025},
  publisher={Taylor \& Francis}
}

@article{ballandies2022decrypting,
  title={Decrypting distributed ledger design—taxonomy, classification and blockchain community evaluation},
  author={Ballandies, Mark C and Dapp, Marcus M and Pournaras, Evangelos},
  journal={Cluster computing},
  volume={25},
  number={3},
  pages={1817--1838},
  year={2022},
  publisher={Springer}
}

@incollection{ballandies2025general,
  title={A General Approach to Tokens and Non-Fungible Tokens},
  author={Ballandies, Mark C and Makode, Parminder Kaur and Tessone, Claudio J},
  booktitle={Routledge Handbook of NFT Law},
  pages={3--16},
  year={2025},
  publisher={Routledge}
}

@inproceedings{ballandies2023taxonomy,
  title={A taxonomy for blockchain-based decentralized physical infrastructure networks (depin)},
  author={Ballandies, Mark C and Wang, Hongyang and Law, Andrew Chung Chee and Yang, Joshua C and G{\"o}sken, Christophe and Andrew, Michael},
  booktitle={2023 IEEE 9th World Forum on Internet of Things (WF-IoT)},
  pages={1--6},
  year={2023},
  organization={IEEE}
}

@article{voshmgir2019foundations,
  title={Foundations of cryptoeconomic systems},
  author={Voshmgir, Shermin and Zargham, Michael and others},
  journal={Research Institute for Cryptoeconomics, Vienna, Working Paper Series/Institute for Cryptoeconomics/Interdisciplinary Research},
  volume={1},
  year={2019}
}

\onecolumn

\appendices
\section{Taxonomy Matrix}
\label{app:rwa-taxonomy-matrix}

\scriptsize
\setlength{\tabcolsep}{3pt}
\renewcommand{\arraystretch}{1.15}

\begin{longtable}{@{}P{2.3cm} P{3.1cm} P{5.1cm} P{6.0cm}@{}}
\caption{RWA taxonomy components, dimensions, definitions, and permitted characteristics.}
\label{tab:rwa-taxonomy-matrix}\\

\toprule
\textbf{Component} & \textbf{Dimension} & \textbf{Definition} & \textbf{Permitted Characteristics} \\
\midrule
\endfirsthead

\toprule
\textbf{Component} & \textbf{Dimension} & \textbf{Definition} & \textbf{Permitted Characteristics} \\
\midrule
\endhead

\midrule
\multicolumn{4}{r}{\textit{Continued on next page}}\\
\endfoot

\bottomrule
\endlastfoot

Governance & Incorporation Jurisdiction
& Country or territory where the issuing legal entity is incorporated or registered.
& Country or territory, e.g., USA, BVI, Cayman Islands, Switzerland. \\

Governance & Legal Entity (Wrapper)
& Legal form through which the tokenized asset is issued.
& Direct issuer; bankruptcy-remote vehicle; SPV; trust; company or equivalent legal form. \\

Governance & Voting
& Whether tokenholders have voting or governance rights in the protocol or issuing entity.
& Yes, with mechanism specified; N/A where no tokenholder voting is identified. \\

Asset & Custody Arrangement
& Arrangement used to safeguard the underlying asset.
& Institutional custodian; issuer self-custody; multi-custodian; decentralized custody; trust arrangement; physical vault storage; N/A. \\

Asset & Ownership Structure
& Legal ownership structure of the underlying asset or claim.
& Direct title; beneficial ownership; SPV-held structure; N/A. \\

Asset & Dispute Forum
& Legal forum or dispute-resolution venue applicable to claims.
& Courts, arbitration forum, governing jurisdiction, or N/A if not disclosed. \\

Asset & Claim Type
& Legal or economic claim represented by the token.
& Direct claim on asset; pro-rata claim on pooled asset; synthetic exposure; debt instrument. \\

Asset & Entitlements
& Actionable rights granted to the holder.
& Redemption right; income distribution right; interest entitlement; governance/voting; usage/access rights; net-profit participation; no additional rights. \\

Asset & Asset Type
& Category of the real-world asset underlying the token.
& Commodity; real estate; fixed income; equities; art and collectibles; carbon credits; other asset class. \\

Asset & Third-Party Audit
& External audit or attestation of the underlying assets or reserves.
& Yes or N/A, with frequency and auditor where disclosed. \\

Asset & On-Chain Proof
& On-chain verification of reserves, asset backing, or pricing.
& N/A; Chainlink Proof of Reserve; smart-contract oracle feed; other proof mechanism. \\

Token Properties & Divisibility
& Whether the token can be held in fractional units.
& Whole-unit only; fractionalized. \\

Token Properties & Settlement Form(s)
& Form of consideration received upon redemption.
& Physical delivery; cash; in-kind token settlement; combination; N/A. \\

Token Properties & Burn
& Whether the token is burned or cancelled upon redemption.
& Yes; N/A. \\

Token Properties & Transferability
& Restrictions on secondary transfers or wallet-to-wallet movement.
& Unrestricted; issuer freeze capability; whitelist-restricted; lock-up period; non-transferable; conditionally restricted. \\

Token Properties & Source of Yield
& Economic source of return for tokenholders.
& Price appreciation only; interest income; rental income; revenue share; staking rewards; lending interest; dividends; coupons; other. \\

Token Properties & Yield Distribution Method
& Mechanism through which yield is delivered to holders.
& Rebase; accrual through NAV or exchange-rate appreciation; distribution; streaming; buyback and burn; yield stripping. \\

Token Properties & Supply
& Whether token supply is fixed or variable.
& Capped, with cap specified where available; uncapped; N/A if not disclosed. \\

DLT & Smart Contract
& Functions automated, secured, or executed through on-chain code.
& Specify implemented functions, e.g., issuance, transfer, redemption, burn, whitelist, oracle, bridge, or N/A. \\

DLT & Chain Deployment
& Blockchain networks on which the token is officially issued or supported.
& Specify supported chain or chains. \\

DLT & DeFi Integrations
& Use of the token in decentralized finance protocols or markets.
& N/A; collateral in lending; liquidity pool; yield vault; structured product; staking; derivatives underlying; other. \\

Economy & Protocol Fees
& Fees charged by the issuer, protocol, fund, or related service providers.
& Minting fee; redemption fee; custody fee; yield spread; transfer tax; management fee; other fee type. \\

Economy & Primary Market Access
& Eligibility requirements for acquiring tokens directly from the issuer.
& Permissionless; KYC-gated; qualified investors only; institutional only; whitelisted only. \\

\end{longtable}

\end{document}